\documentclass[fleqn,11pt,twocolumn]{wlscirep}

\usepackage[utf8]{inputenc}
\usepackage[T1]{fontenc}
\usepackage{siunitx}  
\usepackage[version=4]{mhchem}
\usepackage{float}

\usepackage{soul} 

\definecolor{darkred}{rgb}{0.9,0.0,0}
\usepackage{hyperref}
\hypersetup{
  colorlinks=true,
  urlcolor=darkred,
  linkcolor=darkred,
  citecolor=darkred,
}

\DeclareSIUnit\electron{\mathrm{e^-}}
\DeclareSIUnit\angstrom{\text{Å}}

\title{Mapping Order in Semicrystalline Polymers using
Machine Learning of Nanobeam Electron Diffraction}

\author[1, $\dagger$]{Nicholas Marchese}
\author[1]{Arthur R. C. McCray}
\author[2]{Yael Tsarfati}
\author[3]{Karen Bustillo}
\author[1]{Adam Marks}
\author[1]{Alberto Salleo}
\author[1, $\star$]{Colin Ophus}

\affil[1]{Department of Materials Science and Engineering, Stanford University, 496 Lomita Mall, Stanford, CA 94305, USA}
\affil[2]{SLAC National Accelerator Laboratory, 2575 Sand Hill Road, Menlo Park, CA 94025, USA}
\affil[3]{National Center for Electron Microscopy, Molecular Foundry, Lawrence Berkeley National Laboratory, 1 Cyclotron Road, Berkeley, CA 94720, USA}

\affil[$\dagger$]{njmarch@stanford.edu}
\affil[$\star$]{cophus@stanford.edu}

\begin{abstract}
Organic mixed ionic electronic conductors (OMIECs) are a promising class of polymer materials for applications spanning neuromorphic computation to energy efficient electronics and bioelectronics. Despite being highly tunable, the relationship between structural features and key performance properties such as charge carrier mobility is poorly understood. Scanning nanodiffraction in the transmission electron microscope (TEM) is a powerful probe for elucidating this structure-property relationship, but produces large, noisy datasets that are difficult to interpret because polymer reflections exhibit several distinct morphologies.
To address the complexity, we trained a machine learning (ML) model to detect these polymer diffraction peaks and their intensities from synthetic data. Compared to correlative peak detection algorithms, the conventional method for analyzing nanobeam 4D scanning transmission electron microscopy (4DSTEM) data, we show that the ML model is significantly faster and outperforms correlative algorithms in almost all cases, opening up the possibility of near-live visualization of 4DSTEM experiments.

\end{abstract}

\begin{document}
\maketitle

\section{Introduction}\label{sec1}

Organic mixed ionic electronic conductors (OMIECs) are an emerging class of materials with great promise for neuromorphic computation, energy-efficient electronics, and bioelectronics.\cite{rivnayOrganicElectrochemicalTransistors2018, wongFlexibleElectronicsMaterials2009, paulsenOrganicMixedIonic2020}
They are uniquely suited for these applications due to their ability to conduct both electronic and ionic charges.\cite{vandeburgtNonvolatileOrganicElectrochemical2017, paulsenOrganicMixedIonic2020}
They offer great tunability through synthetic alteration of polymer side chains, polymer backbone engineering, and processing controls to control the microstructure.\cite{mas-torrentRoleMolecularOrder2011, beaujugeMolecularDesignOrdering2011, yaoControlPpStacking2018, paulsenOrganicMixedIonic2020, dongStructureControlPConjugated2019}
However, the relationship between key performance properties such as charge carrier mobility and microstructure is not well understood in OMIECs.\cite{himmelbergerCHARGETRANSPORTSEMICONDUCTING, rivnayOrganicElectrochemicalTransistors2018, paulsenOrganicMixedIonic2020}
Previous studies have relied primarily on X-ray scattering methods, which provide bulk-averaged structural information but do not resolve nanoscale structural heterogeneity.\cite{paulsenOrganicMixedIonic2020, bischakReversibleStructuralPhase2020}

Transmission electron microscopy (TEM) is a powerful method for characterizing the nanoscale structure of polymers, including through phase contrast imaging.\cite{liberaAdvancesTransmissionElectron2010, michlerElectronMicroscopyPolymers2016, cendraUnravelingUnconventionalOrder2021}
Another TEM technique used to characterize the microstructure of polymer systems is four-dimensional scanning TEM (4DSTEM), which records a two-dimensional diffraction pattern at each position in a two-dimensional probe scan.\cite{ophusFourDimensionalScanningTransmission2019, ophusQuantitativeScanningTransmission2023}
4DSTEM has been used to map crystalline orientation of polymers using various diffraction signals like lamellar stacking, backbone periodicity, or $\pi$-$\pi$ stacking.\cite{bustillo4DSTEMBeamSensitiveMaterials2021a}
This technique also enables real-space images to be reconstructed after acquisition using virtual detectors that select specific scattering signals.\cite{bustilloDevelopmentDiffractionScanning2016e, chen2024direct}
It has also been used to study changes in polymer structure with hydration and swelling.\cite{tsarfatiHierarchicalStructureOrganic2025}
Polymer grain morphology and size have been deduced from 4DSTEM datasets through orientation mapping and estimators for persistence length through autocorrelation of diffraction signals.\cite{panovaDiffractionImagingNanocrystalline2019}
All of these methods rely on accurate Bragg peak identification, which remains challenging in polymer materials.

Conjugated polymers often have a semicrystalline microstructure.\cite{uhlmannMicrostructurePolymericMaterials1975}
However, crystallinity in polymers appears differently than in metals, ceramics, or other crystalline materials. 
In previously studied OMIECs, this local order produces three principal diffraction signals associated with lamellar stacking, periodicity along the polymer backbone, and $\pi$-$\pi$ stacking of the aromatic backbone planes.\cite{rivnayQuantitativeDeterminationOrganic2012a}

In addition to the strong $\pi$-$\pi$ reflections found in many conjugated materials, OMIECs such as p(g3T2) produce lamellar and backbone reflections that provide additional structural information but complicate the analysis.
The three reflection classes have distinct morphologies: lamellar stacking produces a sharp pair of reflections near the unscattered beam, backbone periodicity produces sharp arcs alongside other manifestations at slightly larger scattering vector $q$, and $\pi$-$\pi$ stacking produces diffuse intensity distributed over both $q$ and the azimuthal angle $\phi$.
These varied reflection morphologies distinguish polymer diffraction patterns from those of highly crystalline materials, whose more uniform reflections can be identified using cross-correlation templates.
As a result, methods used for Bragg disk detection that were developed for crystalline materials are often neither suitable nor effective for semicrystalline polymer diffraction peak detection.

TEM and 4DSTEM imaging of polymer samples is often limited by their beam sensitivity.\cite{bustilloDevelopmentDiffractionScanning2016e}
To address this, cryogenic TEM---cryogenic cooling of the sample during imaging---is used to substantially reduce beam damage and increase the electron dose before sample degradation.\cite{taylorElectronMicroscopyFrozen, taylorELECTRONDIFFRACTIONFROZEN1974, knapekBeamDamageOrganic1980, donohue2022cryogenic}
Previous work has established detailed methods on how to adopt cryo techniques during 4DSTEM experiments, which greatly improves the resulting dataset quality compared to room-temperature measurements.\cite{bustillo4DSTEMBeamSensitiveMaterials2021a, egertonDelocalizedRadiationDamage2012}
Nevertheless, many polymer systems are both weakly diffracting and highly beam-sensitive; because diffraction data must be acquired at low electron dose to minimize beam damage, the resulting diffraction patterns have a very low signal-to-noise ratio.
Previous studies have used template matching correlation to identify Bragg peak locations, implemented in the py4DSTEM analysis toolkit.\cite{pekinOptimizingDiskRegistration2017, rauchAutomatedNanocrystalOrientation2010, mengImprovementsElectronDiffraction2017, ophusAutomatedCrystalOrientation2022, savitzkyPy4DSTEMSoftwarePackage2021}
More recently, deep-learning frameworks have been developed to automate 4DSTEM preprocessing tasks, including denoising, beam-center calibration, and correction of elliptical distortions.\cite{liu2026unified}
Recent machine-learning approaches have addressed Bragg-disk localization, diffraction-pattern preprocessing, and strain or orientation analysis in 4DSTEM. However, these methods have not targeted the broad, morphologically distinct lamellar, backbone, and $\pi$-$\pi$ reflections observed in semicrystalline OMIECs.\cite{munshiDisentanglingMultipleScattering2022, martineauUnsupervisedMachineLearning2019, yuanTrainingArtificialNeural2021, shiUncoveringMaterialDeformations2022, zintlerMachineLearningAssisted2020}
Additionally, these methods all suffer from both type I errors, where background noise is mistaken for Bragg peaks, and type II errors, where Bragg peak signals are too weak to detect above the background noise.\cite{francis2024clustering}
Since these semicrystalline polymer peaks are so morphologically distinct, tuning the correlative template matching algorithm to perform better on one peak type typically increases the error rates for other peak types.
To precisely map the microstructure of semicrystalline OMIEC systems, we require significantly more accurate and robust diffraction pattern analysis methods.

Here, we present an ML model to provide automated and accurate detection of Bragg peaks for all polymer crystal planes in a semicrystalline sample.
Our model uses a U-Net architecture to accurately determine both the position and intensity of the diffracting signal.
Our method requires only one optionally-user-specified detection threshold parameter, making it substantially more user-friendly and robust to operator bias than previous analytical methods that require more hyperparameters.
For example, the correlation-based py4DSTEM method requires manual tuning of 5 hyperparameters for each peak class, resulting in 15 total hyperparameters to hand tune, potentially per-dataset.
This ML method is also computationally efficient enough to provide a near-real-time map of polymer microstructure to a microscope operator.
By remaining robust to noise in polymer diffraction patterns, this approach enables more complete extraction of information from 4DSTEM datasets without manual peak selection or user-tuned parameters.
This enhanced peak detection will improve orientation mapping, ptychography, polymer domain clustering, grain analysis, virtual image reconstruction, or any future study of polymers with 4DSTEM.

\section{Methods}

\begin{figure}[htbp]
    \centering
    \includegraphics[width=1\linewidth]{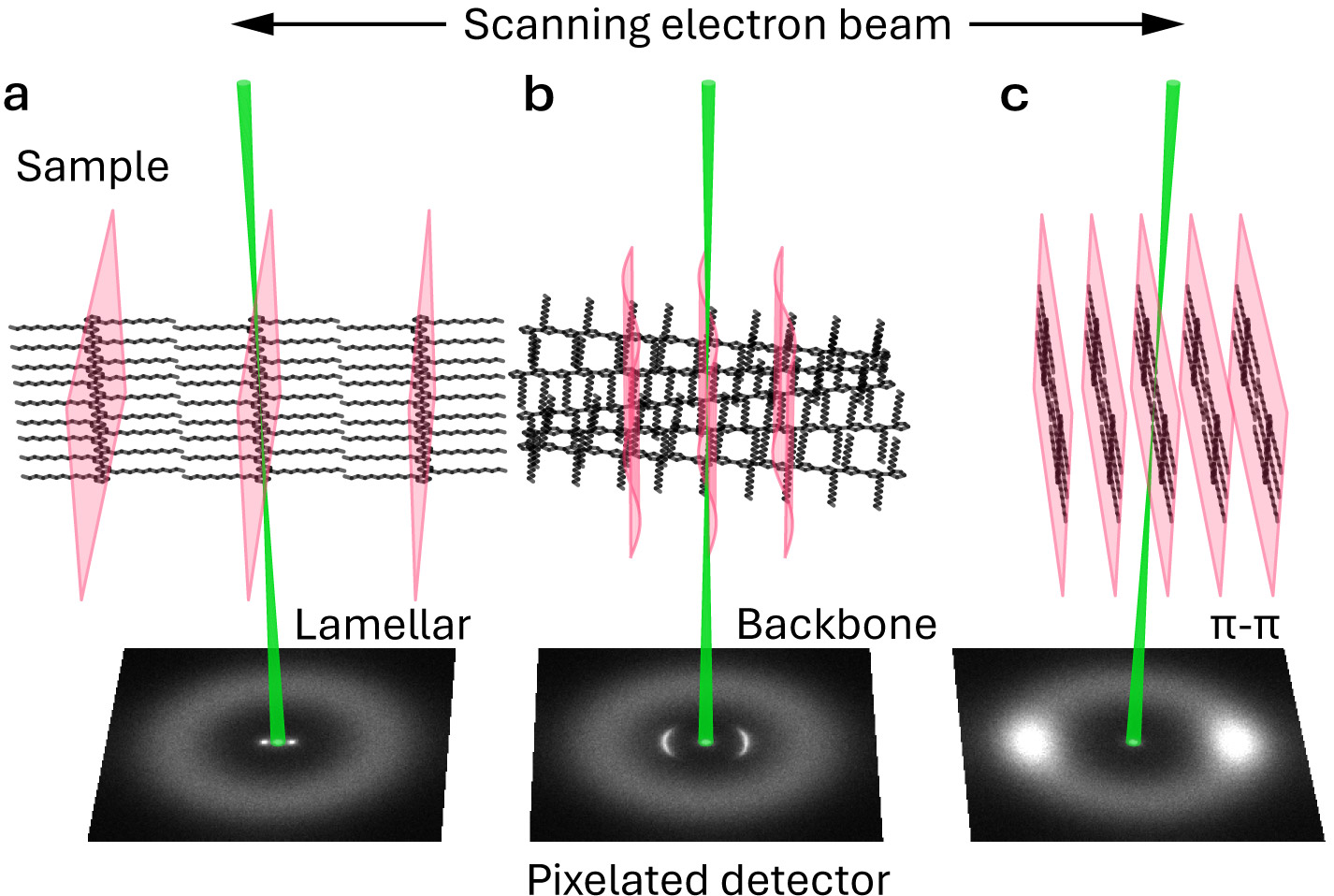}
    \caption{
    \textbf{Schematic of a 4DSTEM experiment on a semicrystalline polymer.} Each panel shows a representative crystallite orientation and the corresponding synthetic diffraction pattern when the indicated Bragg condition is satisfied: (a) lamellar stacking, (b) backbone periodicity, and (c) $\pi$-$\pi$ stacking.
    }
    \label{fig:geometry}
\end{figure}

The 4DSTEM geometry used for this study is shown in Fig.~\ref{fig:geometry}.
Fig.~\ref{fig:geometry}a shows lamellar peaks, which are small disks with low scattering angle due to the high $d$-spacing of the lamellar planes. 
This scattering angle is small enough that the lamellar peaks will typically overlap the central beam. 
High camera lengths are used to minimize this overlap while still retaining view of the other higher-angle diffraction features. 
Fig.~\ref{fig:geometry}b shows backbone peaks, which can have many representations. 
Backbone reflections can appear as radially narrow but azimuthally broad arcs, segmented arcs, isolated peaks, strings of peaks along an arc, or peaks with comet-like tails.
These also commonly present as asymmetric cases, where there is either asymmetry in intensity or presence, breaking 2-fold symmetry.
One example of why this may be the case is when the polymer chain bends into or out of the optic axis, moving into or out of a Bragg condition.
Backbone peaks typically fall between the central beam and the amorphous halo.
Fig.~\ref{fig:geometry}c shows $\pi$-$\pi$ peaks, which are regions of increased intensity sitting atop the amorphous halo demonstrating preferential orientation of the semicrystalline polymer chains due to stacking of the conjugated features (aromatic rings or double-bonded chains) of the polymer backbone.
The signals described above have a variety of annular spreads and intensities, from bright and narrow to dim and broad.
$\pi$-$\pi$ peaks can also demonstrate comet-like trails, where there is a trail of increased intensity occurring clockwise or counter-clockwise from the peak center.

\begin{figure*}[htbp]
    \centering
    \includegraphics[width=1\linewidth]{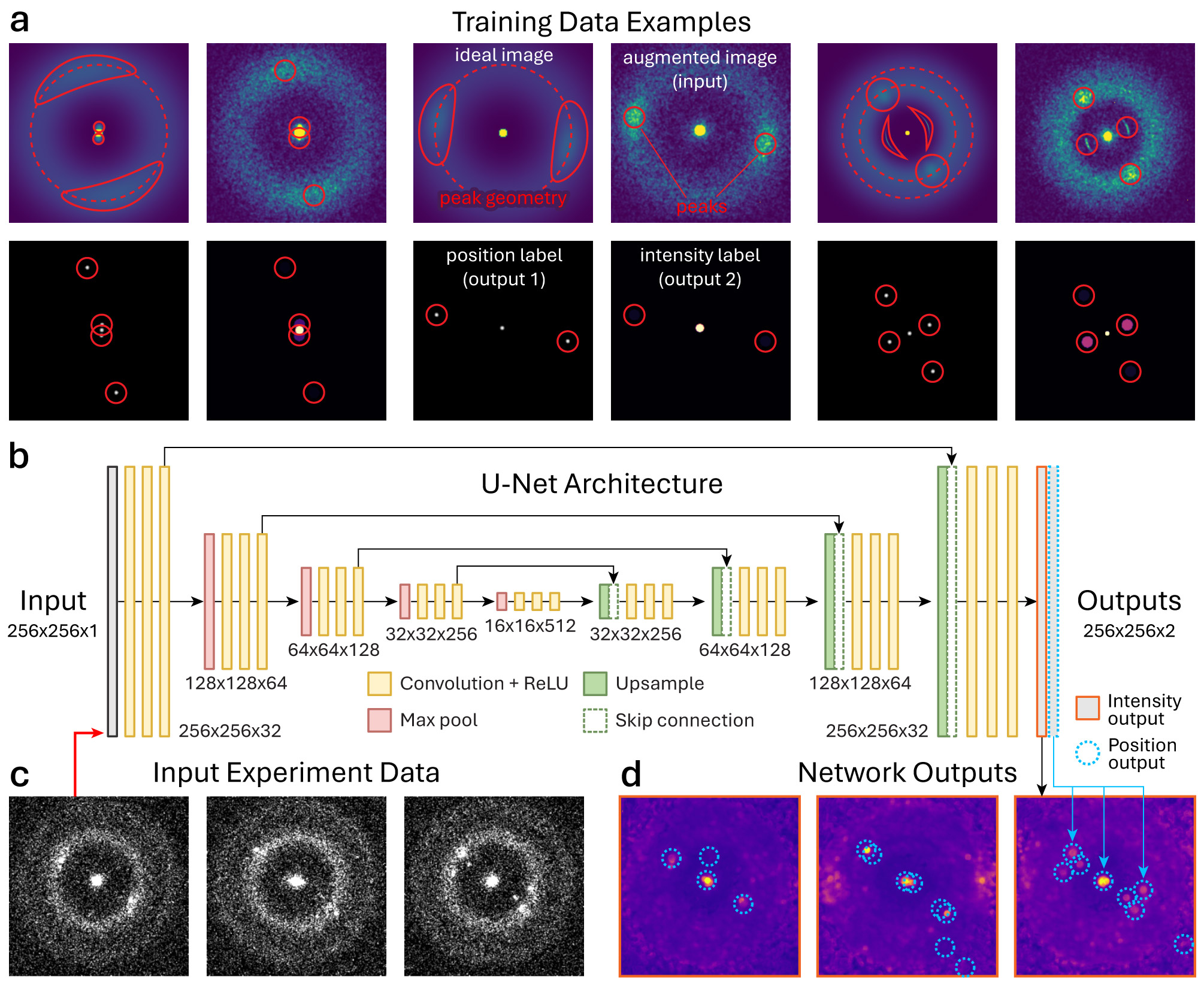}
    \caption{\textbf{Overview of the ML-based peak detection algorithm.} \textbf{(a)} Example synthetic training data used to train the model. The ideal diffraction pattern with relevant geometry is shown and indicated. The augmented diffraction pattern is fed as input to the model. The position and intensity outputs are then compared to the labels. \textbf{(b)} Schematic of the model U-Net architecture. \textbf{(c)} Example experimental input data and \textbf{(d)} associated output position and intensity images. The position image is not shown. The position image maxima, corresponding to detected peaks, are indicated as dotted blue circles on the intensity image. The intensity values are sampled at these positions.}
    \label{fig:overview}
\end{figure*}

Our goal is to train an ML model to detect all crystalline Bragg features in noisy diffraction patterns such as those shown in Fig.~\ref{fig:geometry}a.
Supervised ML requires labeled training data, and direct training on experimental measurements is challenging because the true peak positions and intensities are not known.
We therefore generated 500,000 synthetic data training images.
Three representative synthetic samples are shown in Fig.~\ref{fig:overview}a.
They consist of a ground truth set of peaks, corresponding to some combination of $\pi$-$\pi$, lamellar, and backbone peaks, as well as an amorphous halo at the same scattering angle as the $\pi$-$\pi$ peaks.
Training sets can contain no peaks, single peaks, or multiple sets of peaks.
From each set of peaks, we generate 3 images:
\begin{itemize}
  \item A synthetically generated noisy diffraction pattern.
  \item A peak position label containing sharp 2D Gaussian distributions at all Bragg peak positions.
  \item A peak intensity label containing broad circles with values corresponding to the peak intensity at all Bragg peak positions.
\end{itemize}

We then trained a machine learning model with a U-Net architecture using synthetic diffraction patterns. 
The architecture, visually represented in Fig.~\ref{fig:overview}b, consists of 4 layers with 3 convolutions per layer, 32 starting filters, and a $3\times3$ kernel size. 
The input is a $256\times256$ image with examples shown in Fig.~\ref{fig:overview}c, and the output is two $256\times256$ images with examples shown in Fig.~\ref{fig:overview}d, one channel corresponding to the peak position output, and the other channel corresponding to the intensity output.

We generated the synthetic data using a Python-based simulation workflow.
The generation parameters were initially tuned with reference to experimental data from an oxidized p(g3T2) dataset, an archetypal OMIEC that has been well-studied.\cite{tsarfatiHierarchicalStructureOrganic2025}
We then widely expanded the ranges of the generation parameters to make the model robust when analyzing different types of semicrystalline polymers. 
The pattern generation script first defines different diffraction components, including the central beam, amorphous halo, and various Bragg peak types. 
We then define ``recipes'' in which the different components and their rates (Poisson statistics) are listed. 
Finally, we provide the generator with a list of recipes from which the generator will choose, and the probability weights to choose each recipe. 
The output of the generator is the raw diffraction pattern, a binary image with the peak positions for one label, and the intensities of the peaks for another label, as seen in Fig.~\ref{fig:overview}a. 
To generate the peak-position binary label, we centered a Gaussian peak at each peak coordinate and truncated it with a 3-pixel radial cutoff. 
For the peak intensity label, we centered a circle with a radius of 8 pixels at each peak coordinate with a value corresponding to the intensity of that peak. 
We used a broader intensity label than the peak-position label to make intensity estimation robust to small errors in the predicted peak position.
Overlapping circles for the intensity label are handled according to a Voronoi diagram construction such that the intensity value on the label corresponds to that of the nearest peak.
The central beam footprint is assigned an intensity of 1.0, indicating it is the brightest feature or representing detector saturation, whichever is lower.

We trained the U-Net model on 500,000 diffraction patterns that encompassed the major polymer diffraction components, which are lamellar, backbone, and $\pi$-$\pi$ peaks, as well as other possible peaks such as ice, contamination, or dumbbell peaks.
Generation included the possibility of asymmetry in the peaks for either peak intensity or peak presence, breaking 2-fold symmetry as is sometimes seen experimentally.
The model reached 200 epochs of training, with loss curves presented in Fig. S1.
Model weights at epoch 191 are used as this was the point where training and validation loss were lowest and most closely matched.
To extract peak coordinates and intensities, we applied conventional peak detection to the peak-position output channel and then sampled the peak-intensity output channel at the detected coordinates.
We applied a Gaussian filter to the peak-position output channel for noise suppression. 
We then detected candidate peaks using a local-maximum criterion, requiring each peak pixel to exceed all eight neighboring pixels. We refined the detected coordinates to subpixel accuracy by fitting a quadratic function over a local $3\times3$ pixel neighborhood.
Interpolation of the peak intensity output channel at those coordinates provided the final reported intensity for each peak.

\section{Results and Discussion}

\begin{figure*}[htbp]
    \centering
    \includegraphics[width=1\linewidth]{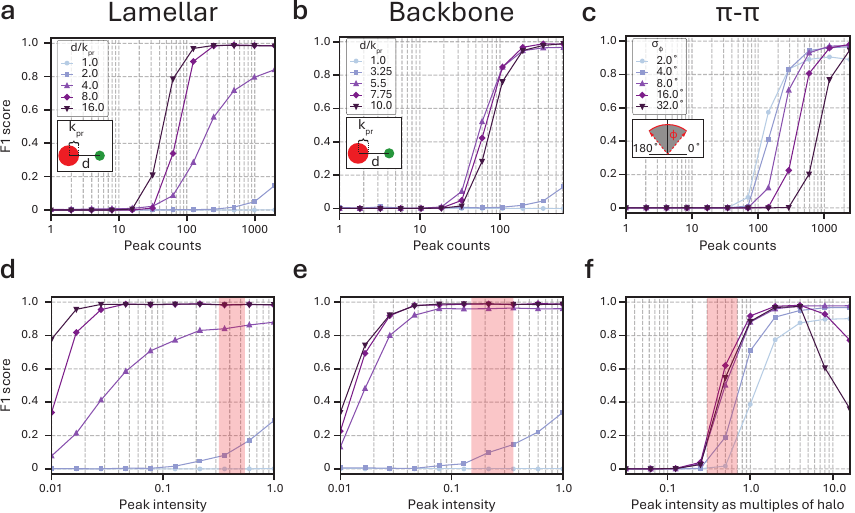}
    \caption{\textbf{Validation results on synthetic data.} Intensity is normalized to either the central beam intensity or detector saturation, whichever is lower. \textbf{(a)--(c)} F1 curves for lamellar, backbone, and $\pi$-$\pi$ peaks, respectively, all with respect to peak counts. \textbf{(d)--(f)} F1 curves for lamellar, backbone, and $\pi$-$\pi$ peaks, respectively, all with respect to peak intensity. Annular spread units, $\sigma_\phi$, are given in degrees. k\textsubscript{pr} is the probe semiangle, and d is the distance from the center of the probe to the center of the peak. d/k\textsubscript{pr} represents the scattering angle of the peak in terms of probe semiangles. The highlighted region indicates estimated intensity ranges seen in the example oxidized p(g3T2) dataset.
    }
    \label{fig:validation}
\end{figure*}

We tested the ML model using synthetic data generated separately from the data used for model training and evaluation. 
We evaluated performance using the F1 score, the harmonic mean of precision and recall, defined as $\frac{2\,TP}{2\,TP + FP + FN}$, where TP, FP, and FN are the numbers of true positives, false positives, and false negatives, respectively.
We evaluated validation performance across varying peak parameters such as intensity, peak radius, and peak annular spread as shown in Fig.~\ref{fig:validation}. 
For each parameter combination, we evaluated 1,000 synthetic diffraction patterns. 
The same 1,000 patterns were reused across all parameter combinations, ensuring that any differences in performance were caused by the varied parameters rather than differences in the test data.
Peak counts were estimated by calculating the expected number of electrons in that peak.
This was done according to:
\begin{equation}
E(c_p) = d_e \cdot (1-w_{bkg}) \cdot \frac{\sum I_p}{\sum I_T}
\end{equation}
where $E(c_p)$ is the expected value of the peak counts, $d_e$ is the total electron dose, $w_{bkg}$ is the background weight fraction, and $\frac{\sum{I_p}}{\sum{I_T}}$ is the peak profile integral over the total image integral of the pre-augmented image without the background. 

For the lamellar stacking peak validation shown in Fig.~\ref{fig:validation}a,d, performance improves monotonically as peak intensity (or peak counts) increases and as peak radial distance increases.
We selected radial distances that begin on the edge of the central beam, where peaks significantly overlap it, and extend outward to larger separations. 
This allowed us to estimate the minimum radial separation required for reliable peak detection.
Peak detection performance of the lamellar peaks starts to increase significantly between a radial distance equal to 2--4 times the probe semiangle.
The performance threshold for lamellar peak detection begins around 50 counts, with the threshold moving to higher counts as the radial distance of the peaks decreases.
Remarkably, even at scattering angles very close to the central beam, to where the peaks partially overlap the central beam, the peak detection of lamellar peaks remains significant at higher intensities.
If d/k\textsubscript{pr}, representing the scattering angle of the peak in terms of probe semiangles, is near 1 this means the center of the peak is on the circumference of the probe, and is partially obscured by the central beam and thus harder to detect.
Additional metrics for lamellar peak validation are shown in Fig. S2 and Fig. S3 with respect to peak counts and peak intensity, respectively.

For the backbone peak validation shown in Fig.~\ref{fig:validation}b,e, performance improves with increasing peak intensity (or peak counts) and increasing radial distance. 
The radial distance for the backbone peaks was evenly distributed across the range chosen, from radial distances smaller than one would expect based on experimental data, to radial distances beyond what would be expected.
The reason the radial distances were distributed evenly is that the backbone peaks show up anywhere between the central beam and the amorphous halo rather than being localized like the lamellar and $\pi$-$\pi$ peaks.
Performance picks up to develop a sharp threshold somewhere between a distance equivalent to 3.25 and 5.5 times the probe semiangle.
The performance threshold for backbone peak detection is around 60--80 counts, with the threshold moving to slightly higher counts as the radial distance of the peak increases.
Additional metrics for backbone peak validation are shown in Fig. S4 and Fig. S5 with respect to peak counts and peak intensity, respectively.

For the $\pi$-$\pi$ stacking peak validation shown in Fig.~\ref{fig:validation}c,f, performance for peak detection initially increases monotonically with peak intensity (or peak counts), and the performance threshold is delayed to higher peak counts as annular spread increases.
Since the $\pi$-$\pi$ peaks are located on the amorphous halo, the annular spread of the peaks is the parameter of interest rather than radial distance. 
The threshold starts around 150 peak counts at the tightest annular spreads with the smallest peak footprints to 800 peak counts at the highest annular spreads with the largest peak footprints.
Given that $\pi$-$\pi$ peaks have the largest peak footprint, significantly larger than either the backbone or lamellar peaks, this reflects how the ML model is able to detect even faint $\pi$-$\pi$ peaks. 
Peak detection across annular spread is generally consistent, spanning from the tightest annular spread where the peak may be viewed more as a disk, to the broader spread notably characteristic of $\pi$-$\pi$ peaks.
However, peak detection for larger annular spreads decreases above a certain intensity threshold, and this degradation begins at lower intensities as the annular spread increases.
This may represent the model struggling to identify the peaks as they begin to spread out to encompass and overshadow the amorphous halo.
Additional metrics for $\pi$-$\pi$ peak validation are shown in Fig. S6 and Fig. S7 with respect to peak counts and peak intensity, respectively.

Notably, the synthetic diffraction patterns used for validation contained only isolated peaks of each type. 
The training dataset used to train the model contained a mixture of patterns with some containing multiple peak types and others containing a single peak type. 
Patterns containing multiple peak classes provide orientational context because lamellar and $\pi$-$\pi$ peaks tend to align, whereas backbone peaks tend to lie approximately perpendicular to them. By testing isolated peak classes, we evaluated the model without these cross-correlations.

\begin{figure*}[htbp]
    \centering
    \includegraphics[width=1\linewidth]{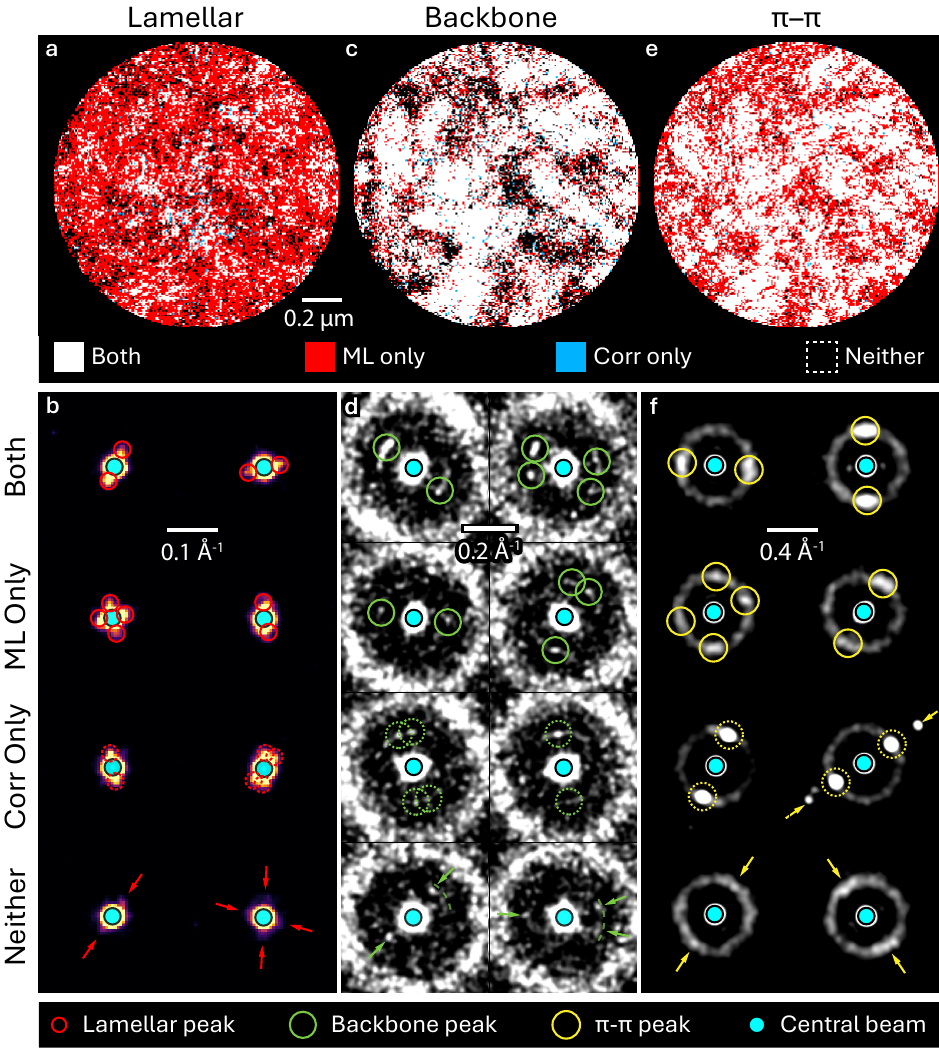}
    \caption{\textbf{Experimental results on oxidized p(g3T2): ML vs. correlative template matching.} Experimental 4DSTEM data from an oxidized p(g3T2) sample were analyzed with both the ML model developed in this work and a correlative peak-detection algorithm, with detected peaks split by peak type for comparison. \textbf{(a), (c), (e)} Difference maps of detected peaks for lamellar, backbone, and $\pi$-$\pi$ peaks, respectively. \textbf{(b), (d), (f)} Representative experimental diffraction patterns for each difference-map category, for lamellar, backbone, and $\pi$-$\pi$ peaks, respectively. Visualization parameters and diffraction-pattern scaling were adjusted per peak type to best show the relevant peaks. Faint or uncertain peaks, including those in the ``neither'' category, were judged by the authors using surrounding probe positions for context.
    }
    \label{fig:exp_detected_map}
\end{figure*}

We next tested the performance of our model on experimental data.
We used our model to analyze peak detection in a 4DSTEM dataset recorded from an oxidized p(g3T2) sample.
We also compared our model against a correlation-based peak-detection algorithm representative of methods used in previous 4DSTEM studies.
We present a comparison of the performance of these two methods in Fig.~\ref{fig:exp_detected_map}. 
The results were split into 3 different types of polymer peaks corresponding to lamellar stacking, backbone periodicity, and $\pi$-$\pi$ stacking. 
Fig.~\ref{fig:exp_detected_map}a, c, and e show peak detection maps for each of those peak types respectively, comparing peak detection for each method.
Fig.~\ref{fig:exp_detected_map}b, d, and f show example experimental diffraction patterns for each of those peak types respectively, highlighting some representative diffraction patterns for each detection case.
The ML model developed in this work outperforms correlative methods in detecting backbone peaks, as seen in Fig.~\ref{fig:exp_detected_map}c, significantly outperforms correlative methods in detecting $\pi$-$\pi$ stacking peaks, as seen in Fig.~\ref{fig:exp_detected_map}e, and greatly outperforms correlative methods in detecting lamellar stacking peaks, as seen in Fig.~\ref{fig:exp_detected_map}a.
The character of the peaks which were detected or missed by each method demonstrates the strengths and weaknesses of the ML model. 

For the lamellar stacking peaks, presented in Fig.~\ref{fig:exp_detected_map}b, the ML model is good at detecting fainter peaks which under inspection have the same orientation as the stronger peaks in neighboring probe positions. 
This implies that the model is more sensitive while also remaining selective, as it can detect fainter peaks which are consistent with surrounding probe positions.
The ML model detects 50,579 lamellar peaks over 26,656 (84.90\%) of the 31,397 total probe positions, while the correlative algorithm detects 7,529 peaks over 7,399 (23.57\%) of the 31,397 total probe positions.
At a small number of probe positions the correlative algorithm detects peaks the ML model did not, but upon inspection these peaks appear very similar to cases where the ML model did detect such peaks in different probe positions.
This may suggest the ML model needs more training data to improve. 
Both methods struggle to detect very weak lamellar peaks.
This may also reflect the desired conservative nature of the model on lamellar peaks, with a balance of sensitivity trading off false negatives in extreme cases for mitigating false positives overall.

For the backbone peaks, presented in Fig.~\ref{fig:exp_detected_map}d, the same pattern holds. 
The ML model detects 49,490 backbone peaks over 24,785 (78.94\%) of the 31,397 total probe positions, while the correlative algorithm detects 23,006 peaks over 20,603 (65.62\%) of the 31,397 total probe positions.
Closer inspection of the experimental data has revealed cases where the model could be made more robust. 
For example, one of the demonstrated backbone peaks shown in Fig.~\ref{fig:exp_detected_map}d, which only the correlative algorithm detected, presents as a very faint, very thin arc with asymmetric intensity. 
Additionally, both methods fail in cases where the peaks are very faint and potentially segmented, which are difficult to identify by eye.
These peaks can still be flagged as likely by eye with additional context of the surrounding probe positions for structural continuity.
This partially reflects the desired conservative nature of the model for backbone peaks, but also elucidates edge cases where the model could be trained on.
Future improvements to encapsulate these complexities will aid in making the ML model more robust.

For the $\pi$-$\pi$ stacking peaks, presented in Fig.~\ref{fig:exp_detected_map}f, the ML model is better than the correlative algorithm at detecting faint peaks. 
The ML model detects 76,815 $\pi$-$\pi$ peaks over 30,082 (95.81\%) of the 31,397 total probe positions, while the correlative algorithm detects 25,569 peaks over 18,853 (60.05\%) of the 31,397 total probe positions.
The ML model outperforms the correlative algorithm in detecting both faint peaks and peaks with intensity asymmetry, as seen in the second row of Fig.~\ref{fig:exp_detected_map}f.
In the cases where the correlative algorithm detected peaks that the ML model did not, the peaks were very strong and sharp, appearing as large and sharp blobs of intensity.
This highlights an edge case a future model could be trained on to make it more robust.
Interestingly, both methods failed to detect a harmonic of one of these strong $\pi$-$\pi$ peaks as shown in the third row, second column of Fig.~\ref{fig:exp_detected_map}f, indicating that harmonics, while uncommon for semicrystalline polymers, should be taken into account in the training data.    
Both models miss peaks that are very faint, to the point of being difficult to determine by eye without additional context from surrounding probe positions.
This again reflects the desired conservative nature of the model.
These results indicate that the ML model appears to be better than the correlative algorithm at detecting faint and asymmetric peaks.

Over the union of lamellar, backbone, and $\pi$-$\pi$ peaks, the ML model detects peaks in 31,288 (99.65\%) of 31,397 total probe positions, while the correlative algorithm detects peaks in 26,737 (85.16\%) of 31,397 total probe positions. 
This indicates the ML model achieves near-complete coverage of the sample area.
The processing time for the ML model on this dataset was 3 minutes and 20 seconds compared to the correlative algorithm's 18 minutes and 21 seconds, translating to the ML model being roughly 5.5$\times$ faster than the correlative algorithm for peak detection when run on a GPU.

\begin{figure*}[htbp]
    \centering
    \includegraphics[width=1\linewidth]{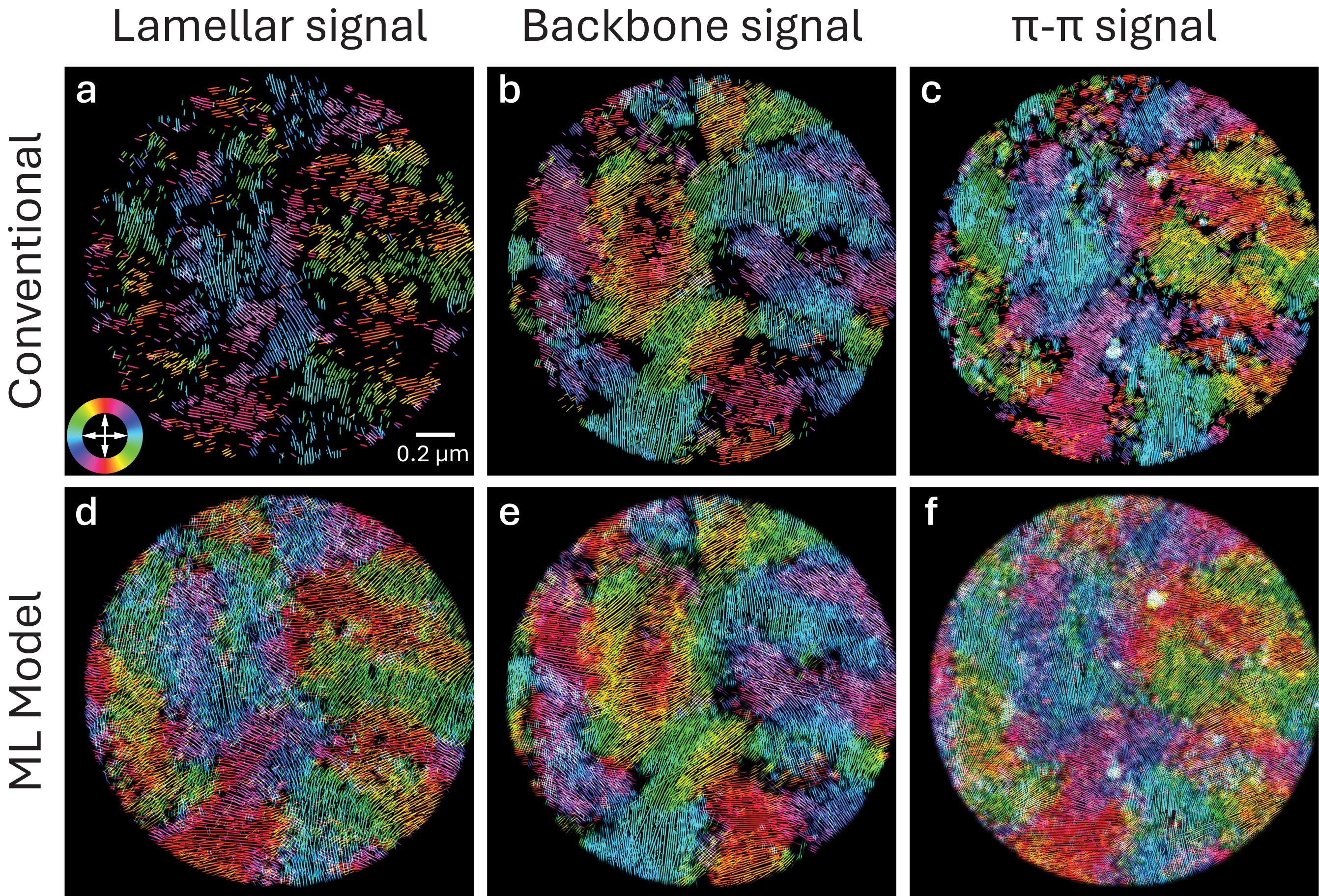}
    \caption{\textbf{Orientation map comparison from experimental oxidized p(g3T2) 4DSTEM data.} \textbf{(a)--(c)} Orientation maps from peaks detected by correlative template matching and \textbf{(d)--(f)} by the ML model developed in this work, for lamellar, backbone, and $\pi$-$\pi$ peaks, respectively. The inset legend indicates directions perpendicular to the crystal planes.
    }
    \label{fig:exp_flowlines}
\end{figure*}

We generated the orientation maps in Fig.~\ref{fig:exp_flowlines} using py4DSTEM\cite{savitzkyPy4DSTEMSoftwarePackage2021} with peaks detected by the ML and conventional correlation-based methods on the same experimental oxidized p(g3T2) dataset.
We generated separate orientation maps for each peak type, where Bragg peaks are drawn as connected lines between adjacent probe positions to show both the local peak orientation and approximate peak intensity, as shown in Fig.~\ref{fig:exp_flowlines}.
These maps provide information on the orientation of the structural features leading to the diffraction conditions observed, and if those peaks are present at those positions on the sample. 
The ML model greatly outperforms the correlative algorithm for lamellar stacking peaks, as seen by the much greater continuity and coverage in the orientation map from the ML model peak data in Fig.~\ref{fig:exp_flowlines}d compared to Fig.~\ref{fig:exp_flowlines}a. 
The ML model also outperforms the correlative algorithm for $\pi$-$\pi$ stacking peaks, as seen in comparing Fig.~\ref{fig:exp_flowlines}f to Fig.~\ref{fig:exp_flowlines}c.
For backbone peaks, the ML model still outperforms correlative algorithms in terms of coverage and continuity, albeit not as drastic, as seen in comparing Fig.~\ref{fig:exp_flowlines}e to Fig.~\ref{fig:exp_flowlines}b.
The density of information captured by the ML model, however, is much greater for the backbone peaks, as well as lamellar and $\pi$-$\pi$ peaks, as shown by the peak dominance maps in Fig. S8.
These results have been inspected by eye to confirm this is not due to false positives but rather genuine peak detection.
This can also be seen reflected in the orientation maps, where for the ML model there are more regions of multiple overlapping lines in different distinct orientations, reflecting the additional peaks the ML model captures that the correlative algorithm did not.
Not only is the coverage greater for the ML model over correlative algorithms, but the ML model also captures more information through more complete peak retrieval than correlative algorithms. Additionally, this model has been tested for generality by running it with 2 additional polymer systems, reduced p(g3T2) and reduced PB2T-TEG. Example diffraction patterns with detected peaks are presented in Fig. S9a,b for reduced p(g3T2) and reduced PB2T-TEG, respectively. Orientation maps are presented in Fig. S10a,b and Fig. S10c,d for reduced p(g3T2) and reduced PB2T-TEG, respectively.

Fig.~\ref{fig:exp_flowlines} emphasizes how the ML model can provide stronger understanding of the orientation fields of these semicrystalline polymers compared to correlative algorithms. 
The ML-based and correlation-based peak detections produce orientation maps with similar overall morphology, providing independent support for the peaks identified by the ML model. 
However, lamellar regions that appeared patchy in the correlation-based analysis are revealed to be nearly continuous in the ML-based analysis.
This indicates that the apparent loss of lamellar order was not necessarily caused by poor crystallinity, poor registry, or weak lamellar stacking. 
Rather, the conventional peak-detection algorithm lacked the sensitivity needed to reveal the underlying continuous crystalline order.

\section{Conclusion}

We have developed a U-Net-based ML model to detect diffraction peaks from noisy 4DSTEM datasets recorded from beam-sensitive polymer samples.
We compared this model to a previously published correlation peak-detection algorithm on both synthetic polymer datasets and experimental oxidized p(g3T2) datasets.
The ML model outperforms the correlative algorithm in nearly every comparison, including processing speed and the elimination of hyperparameters save for a single detection threshold.
A default threshold of 0.5 was sufficient for every case tested, and the ML model ran roughly 5.5$\times$ faster than the correlative algorithm for peak detection on a GPU.

We validated the model against a new set of synthetic data not seen before by the model during either training or evaluation. 
Detection improved as lamellar and backbone peaks moved farther from the central beam. For $\pi$-$\pi$ peaks, detection declined when broad, intense peaks began to dominate the amorphous halo.
Validation also revealed dose thresholds for detection which starts around 60--80 counts for lamellar and backbone peaks, and around 150--800 counts for $\pi$-$\pi$ peaks.
Peak detection almost monotonically increases with intensity, with the exception being the aforementioned inflection point with $\pi$-$\pi$ peaks. 
Comparison of the ML model to correlative methods on an experimental oxidized p(g3T2) dataset reveals additional points for future model improvement, such as increased dataset size, additional peak configurations to improve coverage for how each peak type can manifest, and the addition of peak harmonics.
Orientation maps generated from the experimental oxidized p(g3T2) dataset show that the ML-detected peaks produce more continuous orientational domains than the correlation-based algorithm with greater sample coverage, with the most substantial improvement observed for lamellar peaks.

Our ML model or the approach of using an ML model may be viable for other weakly scattering materials like molecular materials or oxides.
If the material datasets are different enough from those used to train here, the model might require retraining or tuning for those material classes. The model was tested for generality by running it with 3 different polymer material datasets, the oxidized p(g3T2) sample explored here, and separate reduced p(g3T2) and reduced PB2T-TEG datasets which are demonstrated in the supporting information (Fig. S9, S10).

Future work will improve model performance and robustness by refining the synthetic data generation based on the insights from this study, such as including discovered edge cases, adding semicrystalline peak harmonics, and increasing training dataset size. 
Expanded training datasets should better capture missed complexities of experimental diffraction patterns and include additional examples of cases where the current model struggles. 

The current model is already fast enough to support near-real-time analysis and visualization of 4DSTEM scans, where each scan can be processed while the next scan is acquired. 
Interpreting results during microscope operation would allow users to adjust experimental conditions immediately, accelerating experimental iteration.

\section{Data Availability}

The data supporting the findings of this study are available from the corresponding authors upon reasonable request. A public repository link will be added upon publication.

\section{Code Availability}

The code supporting the findings of this study is available from the corresponding authors upon reasonable request. A public repository link will be added upon publication.

\section{Acknowledgments}

This work was primarily supported by the U.S. Department of Energy under Contract No. DE-AC02-76SF00515 through the Basic Energy Sciences (BES) Microelectronics ESTEEM Program (101256). Work at the Molecular Foundry was supported by the Office of Science, Office of Basic Energy Sciences, of the U.S. Department of Energy under Contract No. DE-AC02-05CH11231, under user proposal number(s) MFP-09325. This research used resources of the National Energy Research
Scientific Computing Center, a DOE Office of Science User Facility
supported by the Office of Science of the U.S. Department of Energy
under Contract No. DE-AC02-05CH11231 using NERSC award
NERSC DDR-ERCAP0038484. We thank Karen Ehrhardt and Sangjoon Lee for help with quantEM codebase components used in this project.

\bibliography{paper_1}

\begin{thebibliography}{10}
\urlstyle{rm}
\expandafter\ifx\csname url\endcsname\relax
  \def\url#1{\texttt{#1}}\fi
\expandafter\ifx\csname urlprefix\endcsname\relax\def\urlprefix{URL }\fi
\expandafter\ifx\csname doiprefix\endcsname\relax\def\doiprefix{DOI: }\fi
\providecommand{\bibinfo}[2]{#2}
\providecommand{\eprint}[2][]{\url{#2}}

\bibitem{rivnayOrganicElectrochemicalTransistors2018}
\bibinfo{author}{Rivnay, J.} \emph{et~al.}
\newblock \bibinfo{journal}{\bibinfo{title}{Organic electrochemical
  transistors}}.
\newblock {\emph{\JournalTitle{Nat Rev Mater}}} \textbf{\bibinfo{volume}{3}},
  \bibinfo{pages}{17086} (\bibinfo{year}{2018}).

\bibitem{wongFlexibleElectronicsMaterials2009}
\bibinfo{editor}{Wong, W.~S.} \& \bibinfo{editor}{Salleo, A.} (eds.)
  \emph{\bibinfo{title}{Flexible {{Electronics}}: {{Materials}} and
  {{Applications}}}}, vol.~\bibinfo{volume}{11} of
  \emph{\bibinfo{series}{Electronic {{Materials}}: {{Science}} \&
  {{Technology}}}} (\bibinfo{publisher}{Springer US}, \bibinfo{address}{Boston,
  MA}, \bibinfo{year}{2009}).

\bibitem{paulsenOrganicMixedIonic2020}
\bibinfo{author}{Paulsen, B.~D.}, \bibinfo{author}{Tybrandt, K.},
  \bibinfo{author}{Stavrinidou, E.} \& \bibinfo{author}{Rivnay, J.}
\newblock \bibinfo{journal}{\bibinfo{title}{Organic mixed ionic--electronic
  conductors}}.
\newblock {\emph{\JournalTitle{Nat. Mater.}}} \textbf{\bibinfo{volume}{19}},
  \bibinfo{pages}{13--26} (\bibinfo{year}{2020}).

\bibitem{vandeburgtNonvolatileOrganicElectrochemical2017}
\bibinfo{author}{Van De~Burgt, Y.} \emph{et~al.}
\newblock \bibinfo{journal}{\bibinfo{title}{A non-volatile organic
  electrochemical device as a low-voltage artificial synapse for neuromorphic
  computing}}.
\newblock {\emph{\JournalTitle{Nature Mater}}} \textbf{\bibinfo{volume}{16}},
  \bibinfo{pages}{414--418} (\bibinfo{year}{2017}).

\bibitem{mas-torrentRoleMolecularOrder2011}
\bibinfo{author}{{Mas-Torrent}, M.} \& \bibinfo{author}{Rovira, C.}
\newblock \bibinfo{journal}{\bibinfo{title}{Role of {{Molecular Order}} and
  {{Solid-State Structure}} in {{Organic Field-Effect Transistors}}}}.
\newblock {\emph{\JournalTitle{Chem. Rev.}}} \textbf{\bibinfo{volume}{111}},
  \bibinfo{pages}{4833--4856} (\bibinfo{year}{2011}).

\bibitem{beaujugeMolecularDesignOrdering2011}
\bibinfo{author}{Beaujuge, P.~M.} \& \bibinfo{author}{Fr{\'e}chet, J. M.~J.}
\newblock \bibinfo{journal}{\bibinfo{title}{Molecular {{Design}} and {{Ordering
  Effects}} in {$\pi$}-{{Functional Materials}} for {{Transistor}} and {{Solar
  Cell Applications}}}}.
\newblock {\emph{\JournalTitle{J. Am. Chem. Soc.}}}
  \textbf{\bibinfo{volume}{133}}, \bibinfo{pages}{20009--20029}
  (\bibinfo{year}{2011}).

\bibitem{yaoControlPpStacking2018}
\bibinfo{author}{Yao, Z.-F.}, \bibinfo{author}{Wang, J.-Y.} \&
  \bibinfo{author}{Pei, J.}
\newblock \bibinfo{journal}{\bibinfo{title}{Control of {$\pi$}--{$\pi$}
  {{Stacking}} via {{Crystal Engineering}} in {{Organic Conjugated Small
  Molecule Crystals}}}}.
\newblock {\emph{\JournalTitle{Crystal Growth \& Design}}}
  \textbf{\bibinfo{volume}{18}}, \bibinfo{pages}{7--15} (\bibinfo{year}{2018}).

\bibitem{dongStructureControlPConjugated2019}
\bibinfo{author}{Dong, B.~X.} \emph{et~al.}
\newblock \bibinfo{journal}{\bibinfo{title}{Structure {{Control}} of a
  {$\pi$}-{{Conjugated Oligothiophene-Based Liquid Crystal}} for {{Enhanced
  Mixed Ion}}/{{Electron Transport Characteristics}}}}.
\newblock {\emph{\JournalTitle{ACS Nano}}} \textbf{\bibinfo{volume}{13}},
  \bibinfo{pages}{7665--7675} (\bibinfo{year}{2019}).

\bibitem{himmelbergerCHARGETRANSPORTSEMICONDUCTING}
\bibinfo{author}{Himmelberger, S.}
\newblock \emph{\bibinfo{title}{{{CHARGE TRANSPORT IN SEMICONDUCTING
  POLYMERS}}: {{CONNECTING MICROSTRUCTURE TO ELECTRONIC PROPERTIES}}}}.
\newblock Ph.D. thesis, \bibinfo{school}{Stanford University},
  \bibinfo{address}{Department of Materials Science and Engineering}
  (\bibinfo{year}{2015}).

\bibitem{bischakReversibleStructuralPhase2020}
\bibinfo{author}{Bischak, C.~G.} \emph{et~al.}
\newblock \bibinfo{journal}{\bibinfo{title}{A {{Reversible Structural Phase
  Transition}} by {{Electrochemically-Driven Ion Injection}} into a
  {{Conjugated Polymer}}}}.
\newblock {\emph{\JournalTitle{J. Am. Chem. Soc.}}}
  \textbf{\bibinfo{volume}{142}}, \bibinfo{pages}{7434--7442}
  (\bibinfo{year}{2020}).

\bibitem{liberaAdvancesTransmissionElectron2010}
\bibinfo{author}{Libera, M.~R.} \& \bibinfo{author}{Egerton, R.~F.}
\newblock \bibinfo{journal}{\bibinfo{title}{Advances in the {{Transmission
  Electron Microscopy}} of {{Polymers}}}}.
\newblock {\emph{\JournalTitle{Polymer Reviews}}}
  \textbf{\bibinfo{volume}{50}}, \bibinfo{pages}{321--339}
  (\bibinfo{year}{2010}).

\bibitem{michlerElectronMicroscopyPolymers2016}
\bibinfo{author}{Michler, G.~H.} \& \bibinfo{author}{Lebek, W.}
\newblock \bibinfo{title}{Electron {{Microscopy}} of {{Polymers}}}.
\newblock In \emph{\bibinfo{booktitle}{Polymer {{Morphology}}}},
  chap.~\bibinfo{chapter}{3}, \bibinfo{pages}{37--53},
  \doiprefix\url{10.1002/9781118892756.ch3} (\bibinfo{publisher}{John Wiley \&
  Sons, Ltd}, \bibinfo{year}{2016}).

\bibitem{cendraUnravelingUnconventionalOrder2021}
\bibinfo{author}{Cendra, C.} \emph{et~al.}
\newblock \bibinfo{journal}{\bibinfo{title}{Unraveling the {{Unconventional
  Order}} of a {{High-Mobility Indacenodithiophene}}--{{Benzothiadiazole
  Copolymer}}}}.
\newblock {\emph{\JournalTitle{ACS Macro Lett.}}}
  \textbf{\bibinfo{volume}{10}}, \bibinfo{pages}{1306--1314}
  (\bibinfo{year}{2021}).

\bibitem{ophusFourDimensionalScanningTransmission2019}
\bibinfo{author}{Ophus, C.}
\newblock \bibinfo{journal}{\bibinfo{title}{Four-{{Dimensional Scanning
  Transmission Electron Microscopy}} ({{4D-STEM}}): {{From Scanning
  Nanodiffraction}} to {{Ptychography}} and {{Beyond}}}}.
\newblock {\emph{\JournalTitle{Microsc Microanal}}}
  \textbf{\bibinfo{volume}{25}}, \bibinfo{pages}{563--582}
  (\bibinfo{year}{2019}).

\bibitem{ophusQuantitativeScanningTransmission2023}
\bibinfo{author}{Ophus, C.}
\newblock \bibinfo{journal}{\bibinfo{title}{Quantitative {{Scanning
  Transmission Electron Microscopy}} for {{Materials Science}}: {{Imaging}},
  {{Diffraction}}, {{Spectroscopy}}, and {{Tomography}}}}.
\newblock {\emph{\JournalTitle{Annual Review of Materials Research}}}
  \textbf{\bibinfo{volume}{53}}, \bibinfo{pages}{105--141}
  (\bibinfo{year}{2023}).

\bibitem{bustillo4DSTEMBeamSensitiveMaterials2021a}
\bibinfo{author}{Bustillo, K.~C.} \emph{et~al.}
\newblock \bibinfo{journal}{\bibinfo{title}{{{4D-STEM}} of {{Beam-Sensitive
  Materials}}}}.
\newblock {\emph{\JournalTitle{Acc. Chem. Res.}}}
  \textbf{\bibinfo{volume}{54}}, \bibinfo{pages}{2543--2551}
  (\bibinfo{year}{2021}).

\bibitem{bustilloDevelopmentDiffractionScanning2016e}
\bibinfo{author}{Bustillo, K.~C.} \emph{et~al.}
\newblock \bibinfo{journal}{\bibinfo{title}{Development of {{Diffraction
  Scanning Techniques}} for {{Beam Sensitive Polymers}}.}}
\newblock {\emph{\JournalTitle{Microsc Microanal}}}
  \textbf{\bibinfo{volume}{22}}, \bibinfo{pages}{492--493}
  (\bibinfo{year}{2016}).

\bibitem{chen2024direct}
\bibinfo{author}{Chen, M.} \emph{et~al.}
\newblock \bibinfo{journal}{\bibinfo{title}{Direct imaging of the crystalline
  domains and their orientation in the {PS-b-PEO} block copolymer with
  {4D-STEM}}}.
\newblock {\emph{\JournalTitle{Macromolecules}}} \textbf{\bibinfo{volume}{57}},
  \bibinfo{pages}{5629--5638} (\bibinfo{year}{2024}).

\bibitem{tsarfatiHierarchicalStructureOrganic2025}
\bibinfo{author}{Tsarfati, Y.} \emph{et~al.}
\newblock \bibinfo{journal}{\bibinfo{title}{The hierarchical structure of
  organic mixed ionic--electronic conductors and its evolution in water}}.
\newblock {\emph{\JournalTitle{Nat. Mater.}}} \textbf{\bibinfo{volume}{24}},
  \bibinfo{pages}{101--108} (\bibinfo{year}{2025}).

\bibitem{panovaDiffractionImagingNanocrystalline2019}
\bibinfo{author}{Panova, O.} \emph{et~al.}
\newblock \bibinfo{journal}{\bibinfo{title}{Diffraction imaging of
  nanocrystalline structures in organic semiconductor molecular thin films}}.
\newblock {\emph{\JournalTitle{Nat. Mater.}}} \textbf{\bibinfo{volume}{18}},
  \bibinfo{pages}{860--865} (\bibinfo{year}{2019}).

\bibitem{uhlmannMicrostructurePolymericMaterials1975}
\bibinfo{author}{Uhlmann, D.~R.} \& \bibinfo{author}{Kolbeck, A.~G.}
\newblock \bibinfo{journal}{\bibinfo{title}{The {{Microstructure}} of
  {{Polymeric Materials}}}}.
\newblock {\emph{\JournalTitle{Scientific American}}}
  \textbf{\bibinfo{volume}{233}}, \bibinfo{pages}{96--107}
  (\bibinfo{year}{1975}).
\newblock \eprint{24949965}.

\bibitem{rivnayQuantitativeDeterminationOrganic2012a}
\bibinfo{author}{Rivnay, J.}, \bibinfo{author}{Mannsfeld, S. C.~B.},
  \bibinfo{author}{Miller, C.~E.}, \bibinfo{author}{Salleo, A.} \&
  \bibinfo{author}{Toney, M.~F.}
\newblock \bibinfo{journal}{\bibinfo{title}{Quantitative {{Determination}} of
  {{Organic Semiconductor Microstructure}} from the {{Molecular}} to {{Device
  Scale}}}}.
\newblock {\emph{\JournalTitle{Chem. Rev.}}} \textbf{\bibinfo{volume}{112}},
  \bibinfo{pages}{5488--5519} (\bibinfo{year}{2012}).

\bibitem{taylorElectronMicroscopyFrozen}
\bibinfo{author}{Taylor, K.~A.} \& \bibinfo{author}{Glaeser, R.~M.}
\newblock \bibinfo{journal}{\bibinfo{title}{Electron {{Microscopy}} of {{Frozen
  Hydrated Biological Specimens}}}}.
\newblock {\emph{\JournalTitle{Journal of Ultrastructure Research}}}
  \textbf{\bibinfo{volume}{55}}, \bibinfo{pages}{448--456}
  (\bibinfo{year}{1976}).

\bibitem{taylorELECTRONDIFFRACTIONFROZEN1974}
\bibinfo{author}{Taylor, K.~A.} \& \bibinfo{author}{Glaeser, R.~M.}
\newblock \bibinfo{journal}{\bibinfo{title}{{{ELECTRON DIFFRACTION OF FROZEN}},
  {{HYDRATED PROTEIN CRYSTALS}}}}.
\newblock {\emph{\JournalTitle{Science}}}  (\bibinfo{year}{1974}).

\bibitem{knapekBeamDamageOrganic1980}
\bibinfo{author}{Knapek, E.} \& \bibinfo{author}{Dubochet, J.}
\newblock \bibinfo{journal}{\bibinfo{title}{Beam damage to organic material is
  considerably reduced in cryo-electron microscopy}}.
\newblock {\emph{\JournalTitle{Journal of Molecular Biology}}}
  \textbf{\bibinfo{volume}{141}}, \bibinfo{pages}{147--161}
  (\bibinfo{year}{1980}).

\bibitem{donohue2022cryogenic}
\bibinfo{author}{Donohue, J.} \emph{et~al.}
\newblock \bibinfo{journal}{\bibinfo{title}{Cryogenic {4D-STEM} analysis of an
  amorphous-crystalline polymer blend: Combined nanocrystalline and amorphous
  phase mapping}}.
\newblock {\emph{\JournalTitle{{iScience}}}} \textbf{\bibinfo{volume}{25}}
  (\bibinfo{year}{2022}).

\bibitem{egertonDelocalizedRadiationDamage2012}
\bibinfo{author}{Egerton, R.~F.}, \bibinfo{author}{Lazar, S.} \&
  \bibinfo{author}{Libera, M.}
\newblock \bibinfo{journal}{\bibinfo{title}{Delocalized radiation damage in
  polymers}}.
\newblock {\emph{\JournalTitle{Micron}}} \textbf{\bibinfo{volume}{43}},
  \bibinfo{pages}{2--7} (\bibinfo{year}{2012}).

\bibitem{pekinOptimizingDiskRegistration2017}
\bibinfo{author}{Pekin, T.~C.}, \bibinfo{author}{Gammer, C.},
  \bibinfo{author}{Ciston, J.}, \bibinfo{author}{Minor, A.~M.} \&
  \bibinfo{author}{Ophus, C.}
\newblock \bibinfo{journal}{\bibinfo{title}{Optimizing disk registration
  algorithms for nanobeam electron diffraction strain mapping}}.
\newblock {\emph{\JournalTitle{Ultramicroscopy}}}
  \textbf{\bibinfo{volume}{176}}, \bibinfo{pages}{170--176}
  (\bibinfo{year}{2017}).

\bibitem{rauchAutomatedNanocrystalOrientation2010}
\bibinfo{author}{Rauch, E.~F.} \emph{et~al.}
\newblock \bibinfo{journal}{\bibinfo{title}{Automated nanocrystal orientation
  and phase mapping in the transmission electron microscope on the basis of
  precession electron diffraction}}.
\newblock {\emph{\JournalTitle{Zeitschrift f\"ur Kristallographie}}}
  \textbf{\bibinfo{volume}{225}}, \bibinfo{pages}{103--109}
  (\bibinfo{year}{2010}).

\bibitem{mengImprovementsElectronDiffraction2017}
\bibinfo{author}{Meng, Y.} \& \bibinfo{author}{Zuo, J.-M.}
\newblock \bibinfo{journal}{\bibinfo{title}{Improvements in {{Electron
  Diffraction Pattern Automatic Indexing Algorithms}}}}.
\newblock {\emph{\JournalTitle{The European Physical Journal Applied Physics}}}
  \textbf{\bibinfo{volume}{80}}, \bibinfo{pages}{10701} (\bibinfo{year}{2017}).

\bibitem{ophusAutomatedCrystalOrientation2022}
\bibinfo{author}{Ophus, C.} \emph{et~al.}
\newblock \bibinfo{journal}{\bibinfo{title}{Automated {{Crystal Orientation
  Mapping}} in {{py4DSTEM}} using {{Sparse Correlation Matching}}}}.
\newblock {\emph{\JournalTitle{Microanal}}} \textbf{\bibinfo{volume}{28}},
  \bibinfo{pages}{390--403} (\bibinfo{year}{2022}).

\bibitem{savitzkyPy4DSTEMSoftwarePackage2021}
\bibinfo{author}{Savitzky, B.~H.} \emph{et~al.}
\newblock \bibinfo{journal}{\bibinfo{title}{{{py4DSTEM}}: {{A Software
  Package}} for {{Four-Dimensional Scanning Transmission Electron Microscopy
  Data Analysis}}}}.
\newblock {\emph{\JournalTitle{Microanal}}} \textbf{\bibinfo{volume}{27}},
  \bibinfo{pages}{712--743} (\bibinfo{year}{2021}).

\bibitem{liu2026unified}
\bibinfo{author}{Liu, M.} \emph{et~al.}
\newblock \bibinfo{journal}{\bibinfo{title}{A unified preprocessing framework
  for high-throughput diffraction pattern analysis}}.
\newblock {\emph{\JournalTitle{npj Computational Materials}}}
  \textbf{\bibinfo{volume}{12}}, \bibinfo{pages}{145} (\bibinfo{year}{2026}).

\bibitem{munshiDisentanglingMultipleScattering2022}
\bibinfo{author}{Munshi, J.} \emph{et~al.}
\newblock \bibinfo{journal}{\bibinfo{title}{Disentangling multiple scattering
  with deep learning: Application to strain mapping from electron diffraction
  patterns}}.
\newblock {\emph{\JournalTitle{npj Comput Mater}}}
  \textbf{\bibinfo{volume}{8}}, \bibinfo{pages}{254} (\bibinfo{year}{2022}).

\bibitem{martineauUnsupervisedMachineLearning2019}
\bibinfo{author}{Martineau, B.~H.}, \bibinfo{author}{Johnstone, D.~N.},
  \bibinfo{author}{{van Helvoort}, A. T.~J.}, \bibinfo{author}{Midgley, P.~A.}
  \& \bibinfo{author}{Eggeman, A.~S.}
\newblock \bibinfo{journal}{\bibinfo{title}{Unsupervised machine learning
  applied to scanning precession electron diffraction data}}.
\newblock {\emph{\JournalTitle{Adv Struct Chem Imag}}}
  \textbf{\bibinfo{volume}{5}}, \bibinfo{pages}{3} (\bibinfo{year}{2019}).

\bibitem{yuanTrainingArtificialNeural2021}
\bibinfo{author}{Yuan, R.}, \bibinfo{author}{Zhang, J.}, \bibinfo{author}{He,
  L.} \& \bibinfo{author}{Zuo, J.-M.}
\newblock \bibinfo{journal}{\bibinfo{title}{Training artificial neural networks
  for precision orientation and strain mapping using {{4D}} electron
  diffraction datasets}}.
\newblock {\emph{\JournalTitle{Ultramicroscopy}}}
  \textbf{\bibinfo{volume}{231}}, \bibinfo{pages}{113256}
  (\bibinfo{year}{2021}).

\bibitem{shiUncoveringMaterialDeformations2022}
\bibinfo{author}{Shi, C.} \emph{et~al.}
\newblock \bibinfo{journal}{\bibinfo{title}{Uncovering material deformations
  via machine learning combined with four-dimensional scanning transmission
  electron microscopy}}.
\newblock {\emph{\JournalTitle{npj Comput Mater}}}
  \textbf{\bibinfo{volume}{8}}, \bibinfo{pages}{114} (\bibinfo{year}{2022}).

\bibitem{zintlerMachineLearningAssisted2020}
\bibinfo{author}{Zintler, A.} \emph{et~al.}
\newblock \bibinfo{journal}{\bibinfo{title}{Machine {{Learning Assisted Pattern
  Matching}}: {{Insight}} into {{Oxide Electronic Device Performance}} by
  {{Phase Determination}} in {{4D-STEM Datasets}}}}.
\newblock {\emph{\JournalTitle{Microsc Microanal}}}
  \textbf{\bibinfo{volume}{26}}, \bibinfo{pages}{1908--1909}
  (\bibinfo{year}{2020}).

\bibitem{francis2024clustering}
\bibinfo{author}{Francis, C.} \& \bibinfo{author}{Voyles, P.~M.}
\newblock \bibinfo{journal}{\bibinfo{title}{Clustering characteristic
  diffraction vectors in {4-D STEM} data sets from overlapping structures in
  nanocrystalline and amorphous materials}}.
\newblock {\emph{\JournalTitle{Ultramicroscopy}}}
  \textbf{\bibinfo{volume}{267}}, \bibinfo{pages}{114040}
  (\bibinfo{year}{2024}).

\end{thebibliography}

\clearpage
\onecolumn                                    %
\section*{Supporting Information}

\setcounter{figure}{0}
\renewcommand{\thefigure}{S\arabic{figure}}

\begin{figure*}[htbp]
    \centering
    \includegraphics[width=1\linewidth]{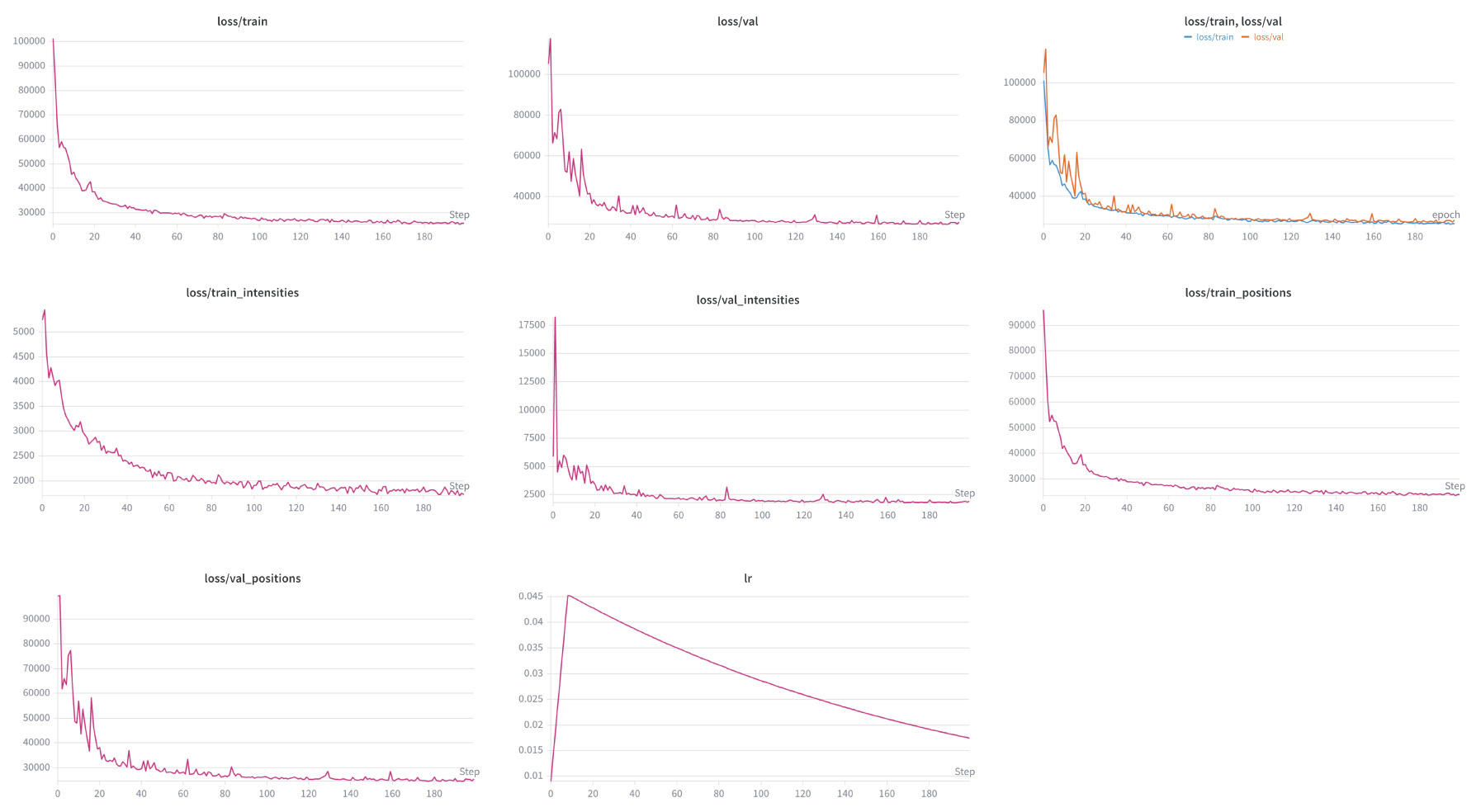}
    \caption{\textbf{Various loss and parameter curves from the ML training run for the model developed in this work.}}
    \label{fig:loss_curves}
\end{figure*}

\begin{figure*}[htbp]
    \centering
    \includegraphics[width=1\linewidth]{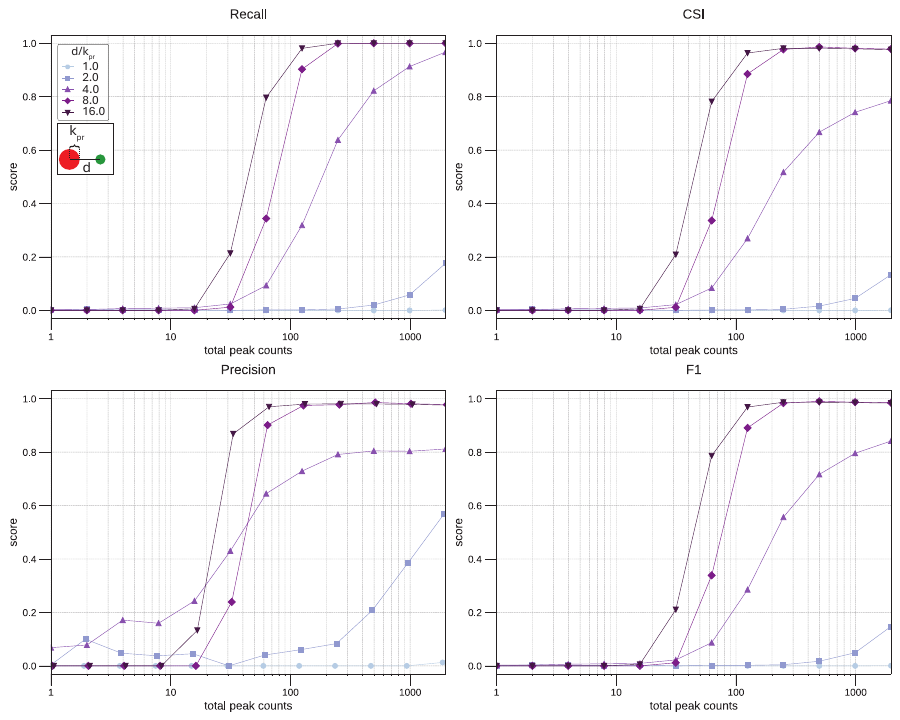}
    \caption{\textbf{Various validation statistics for lamellar peaks with respect to counts.}}
    \label{fig:lamellar_lines_counts}
\end{figure*}

\begin{figure*}[htbp]
    \centering
    \includegraphics[width=1\linewidth]{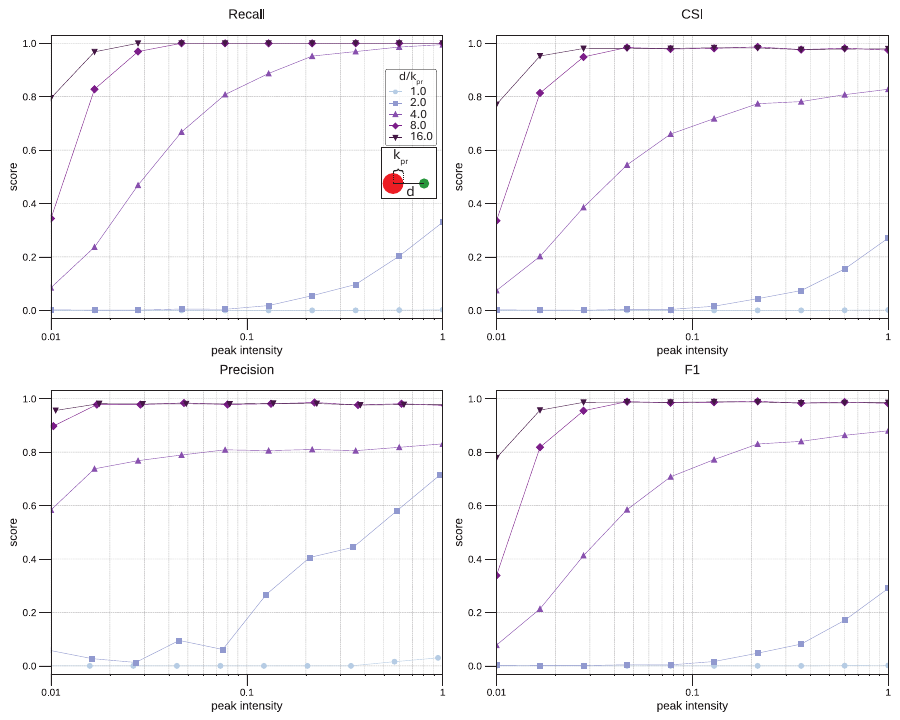}
    \caption{\textbf{Various validation statistics for lamellar peaks with respect to intensity.}}
    \label{fig:lamellar_lines_intensity}
\end{figure*}

\begin{figure*}[htbp]
    \centering
    \includegraphics[width=1\linewidth]{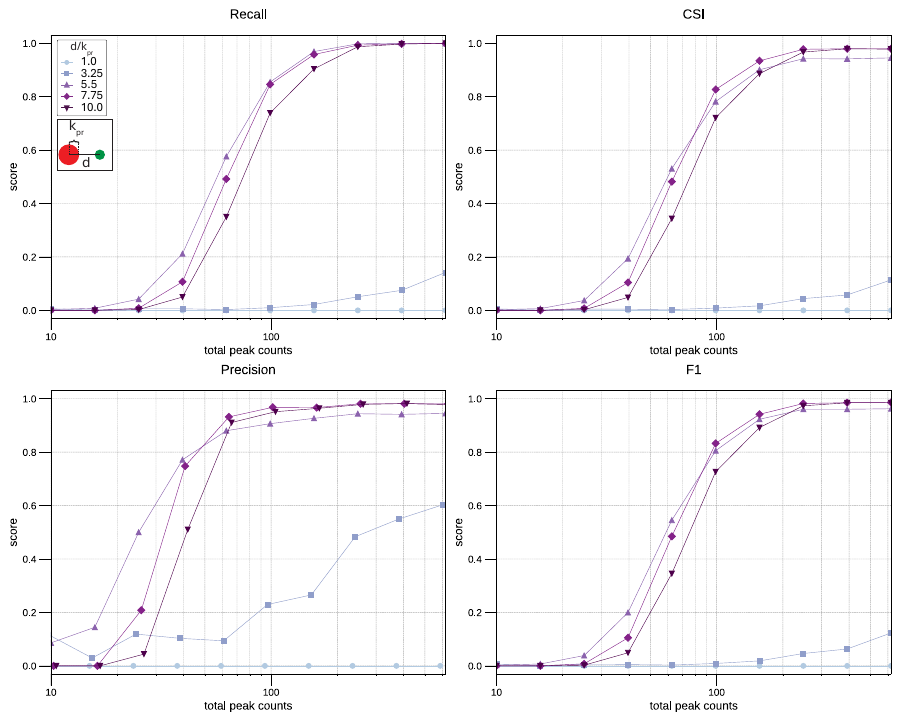}
    \caption{\textbf{Various validation statistics for backbone peaks with respect to counts.}}
    \label{fig:backbone_lines_counts}
\end{figure*}

\begin{figure*}[htbp]
    \centering
    \includegraphics[width=1\linewidth]{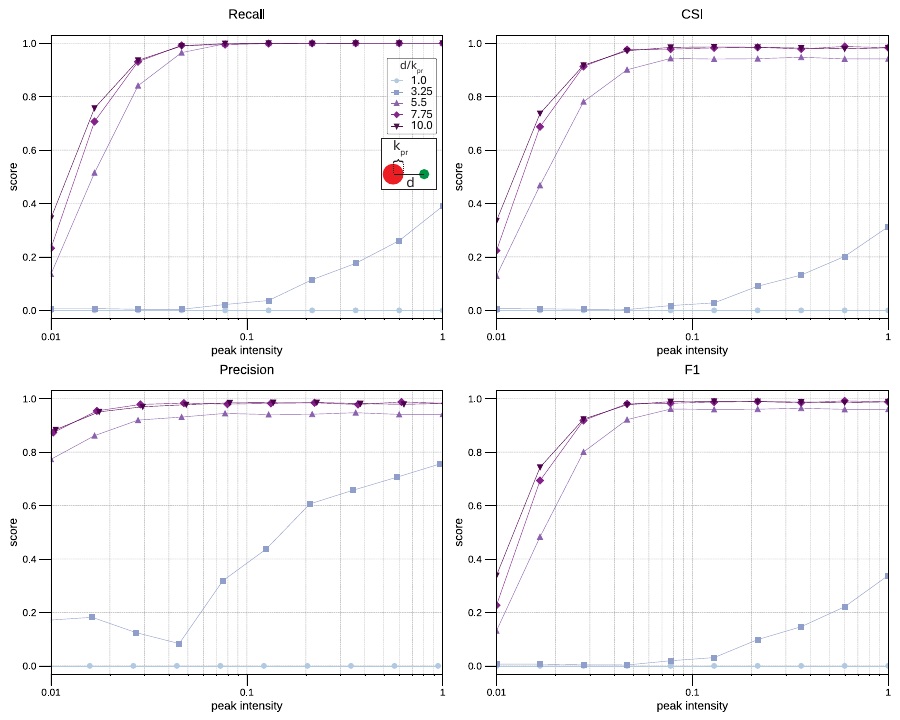}
    \caption{\textbf{Various validation statistics for backbone peaks with respect to intensity.}}
    \label{fig:backbone_lines_intensity}
\end{figure*}

\begin{figure*}[htbp]
    \centering
    \includegraphics[width=1\linewidth]{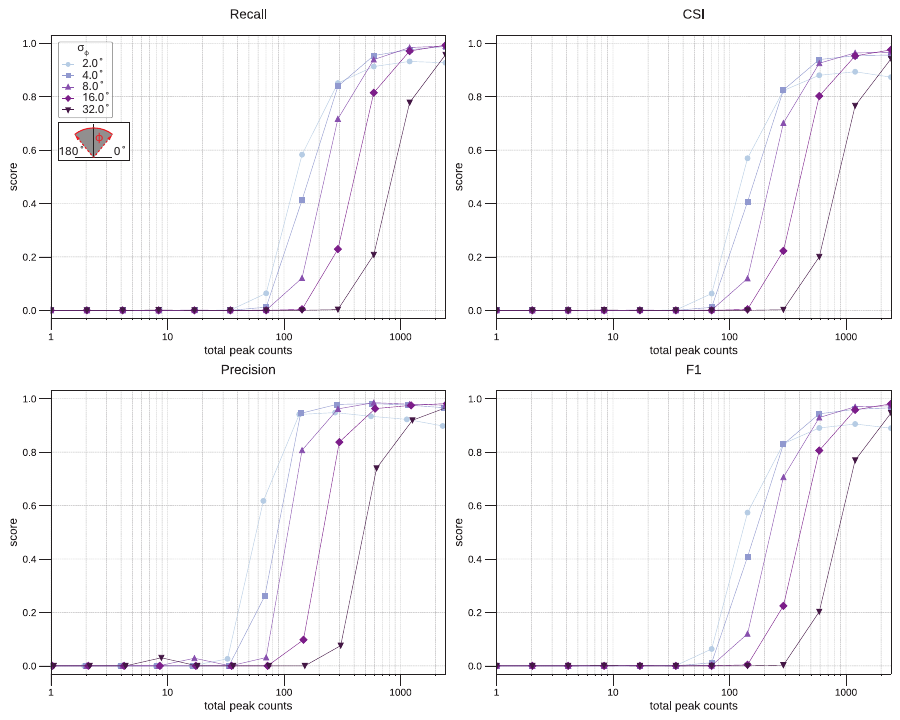}
    \caption{\textbf{Various validation statistics for $\pi$-$\pi$ peaks with respect to counts.}}
    \label{fig:pipi_lines_counts}
\end{figure*}

\begin{figure*}[htbp]
    \centering
    \includegraphics[width=1\linewidth]{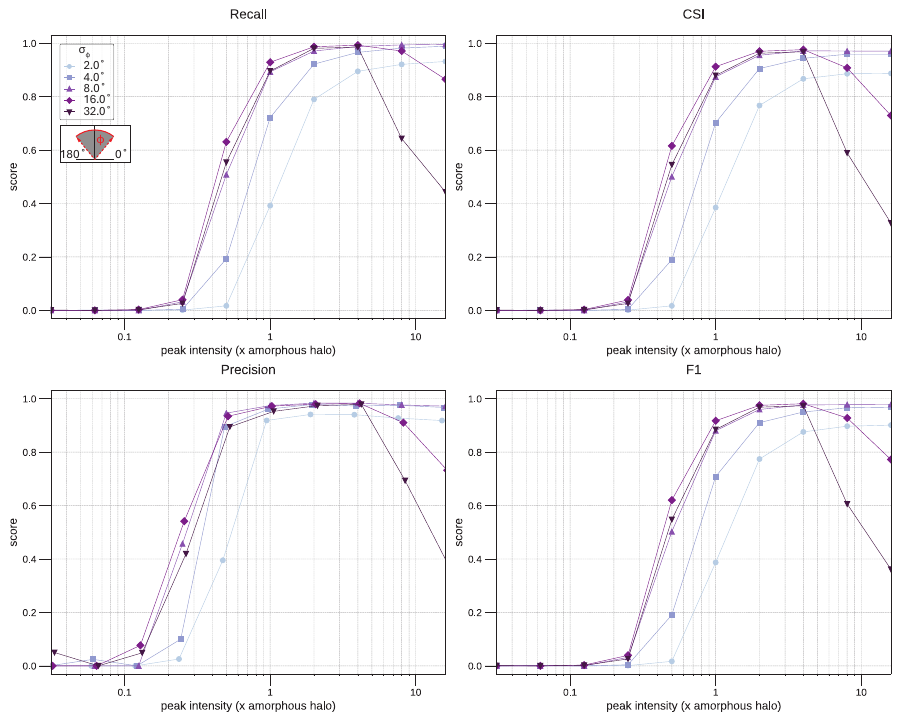}
    \caption{\textbf{Various validation statistics for $\pi$-$\pi$ peaks with respect to intensity in terms of multiples of amorphous halo intensity.}}
    \label{fig:pipi_lines_intensity}
\end{figure*}

\begin{figure*}[htbp]
    \centering
    \includegraphics[width=1\linewidth]{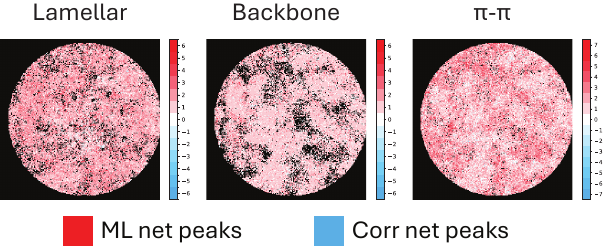}
    \caption{\textbf{Peak dominance maps for an experimental p(g3T2) dataset comparing results from the ML model developed in this work to a conventional correlative template matching algorithm.} The color indicates how many more peaks were detected by one method over the other, where red indicates more peaks detected by the ML model, blue more peaks detected by the correlative algorithm, and the intensity of the color indicating how many peaks.}
    \label{fig:peak_dominance_maps}
\end{figure*}

\begin{figure*}[htbp]
    \centering
    \includegraphics[width=1\linewidth]{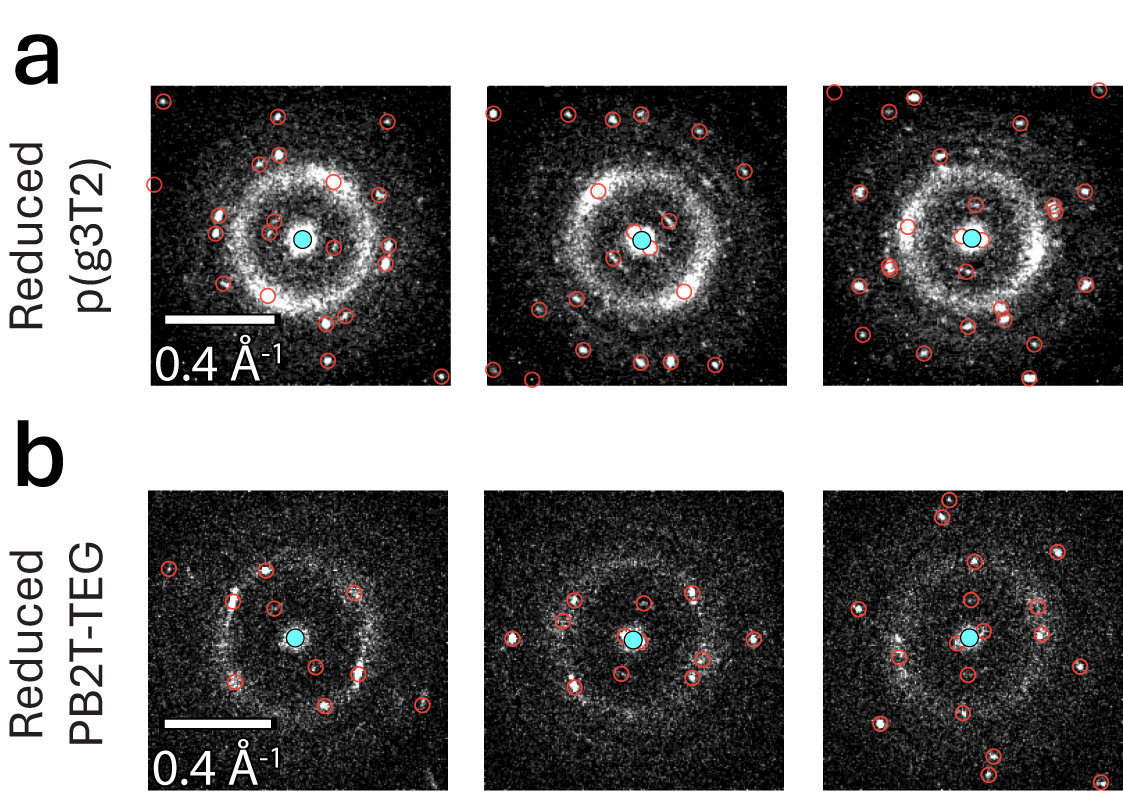}
    \caption{\textbf{ML model peak detection on reduced p(g3T2) and PB2T-TEG polymer diffraction patterns.} Representative diffraction patterns with overlaid peaks detected by the ML model for \textbf{(a)} a reduced p(g3T2) sample and \textbf{(b)} a reduced PB2T-TEG sample with known pinholes.}
    \label{fig:pb2tteg_p3ht_diffraction_patterns}
\end{figure*}

\begin{figure*}[htbp]
    \centering
    \includegraphics[width=1\linewidth]{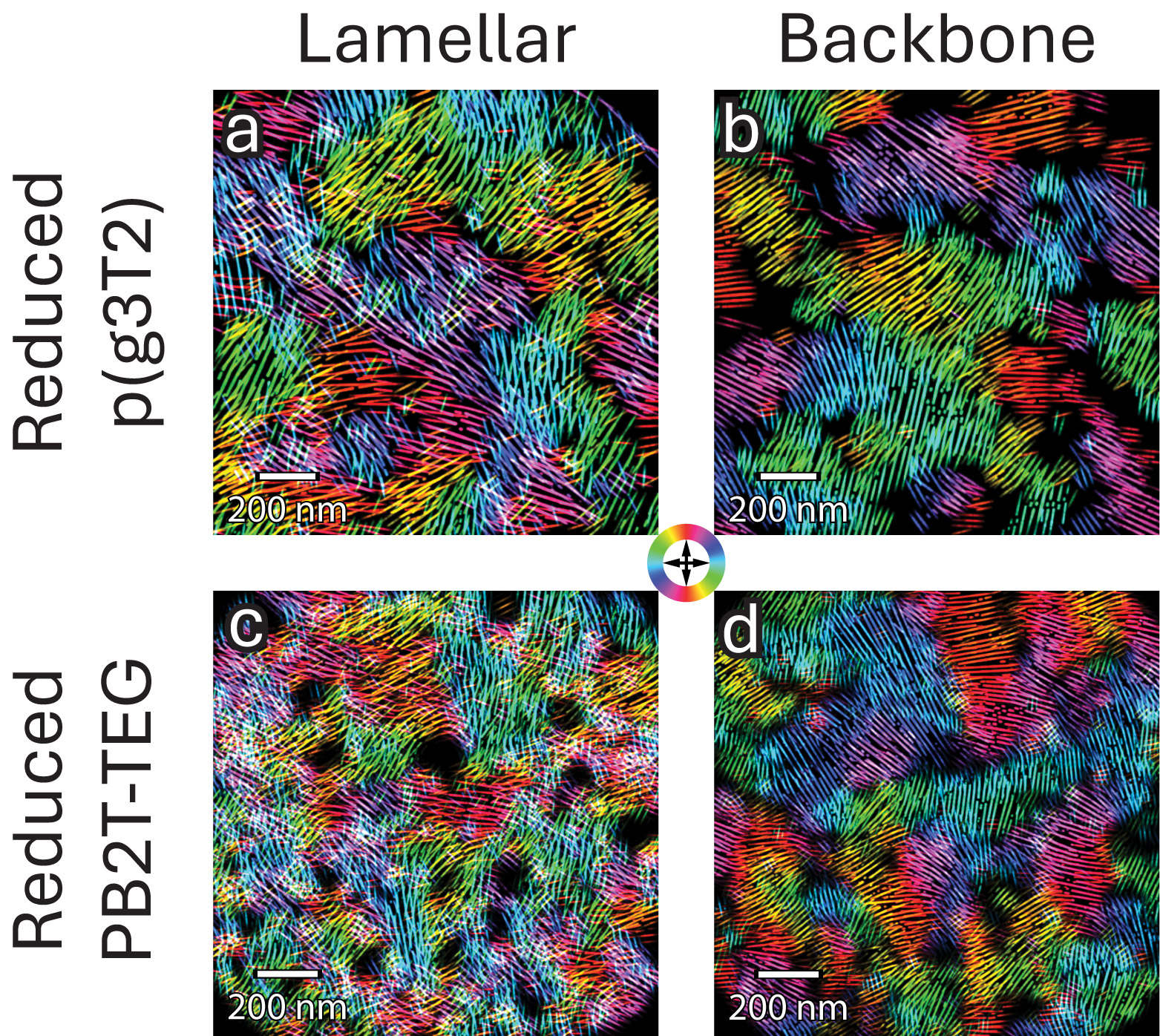}
    \caption{\textbf{Orientation maps for reduced p(g3T2) and PB2T-TEG from ML model detected peaks.} Orientation maps from peaks detected by the ML model developed in this work for \textbf{(a), (b)} a reduced p(g3T2) sample for lamellar and backbone peaks respectively and \textbf{(c), (d)} a reduced PB2T-TEG sample with known pinholes for lamellar and backbone peaks respectively. The inset legend indicates directions perpendicular to the crystal planes.}
    \label{fig:pb2tteg_p3ht_flowlines}
\end{figure*}

\end{document}